\documentclass[12pt]{article}
\usepackage{latexsym}
\usepackage{graphicx}
\begin{document}

\begin{center}
{\bf \LARGE Parameterized total cross sections for  \\ \vspace*{.4cm} pion production in nuclear collisions}
\end{center}

\begin{center}
John W. Norbury $^{\rm a, \star}$,  Lawrence W. Townsend $^{\rm b}$ \\
\end{center}

\begin{center}
{\small  {\em  $^{\rm a}$ University of Southern Mississippi, Hattiesburg, MS, 39406, USA \\

$^{\rm b}$ University of Tennessee, Knoxville, TN, 37996-2300,  USA}}
 \end{center}

\begin{center}
 {\em email:} john.norbury@usm.edu, 
ltownsen@tennessee.edu
\end{center}

\noindent \hrulefill

\noindent {\bf Abstract}\\

Total inclusive cross sections for neutral and charged pion production in proton-nucleus and nucleus-nucleus reactions have been calculated and compared to experiment.    Nucleon-nucleon theoretical cross sections have been scaled up to nuclear collisions using a scaling factor similar to $(A_PA_T)^{2/3}$, where $A_P$ and $A_T$ are the nucleon numbers of the projectile and target nuclei. Variations in the power of this scaling factor have been studied and a good  fit to experiment is obtained with a small  modification of the power.  Theoretical cross sections are written in a form  that is very suitable for immediate input into   transport codes.\\



\noindent \hrulefill

\section{Introduction}

Radiation transport codes \cite{rp} often use arithmetic parameterizations of cross sections, in order that the transport code is able to run quickly. We have previously developed parameterizations of pion production cross sections in proton-proton reactions \cite{blattnig}. It is worthwhile  to extend these to proton-nucleus and nucleus-nucleus reactions for use in heavy-ion transport codes, which are needed for space radiation protection \cite{rp}  and other applications.
Pion production is important in several ways. After neutral pions are produced they decay electromagnetically to high energy photons which initiate an electromagnetic cascade consisting of other high energy photons, electrons and positrons. The high energy photons, resulting from nucleon-nucleon collisions in the interstellar medium, are easily seen by gamma ray telescopes. The high energy electrons can penetrate space suits and therefore are important during EVA (extra-vehicular activity). Charged pions can produce ionization before decaying and are therefore a source of radiation by themselves. High energy muons result from their decay and it is well known that these muons can reach the surface of the Earth due to relativistic time dilation. Such muons can also penetrate space suits. The peak of the cosmic ray spectrum occurs near  the GeV region, which we refer to as intermediate energy. This is the region of interest for space radiation applications.
 The aim of the present work is to provide simple parameterizations of total pion production cross sections in nuclear collisions, which can be used as input for heavy-ion transport codes.

\section{Nucleon-nucleon  cross sections}
The total  inclusive cross section for pion production in proton-proton collisions has been parameterized as \cite{blattnig}
\begin{eqnarray}
\sigma_{pp \rightarrow \pi^0 X} &=& \left ( 0.007+0.1 \frac{{\rm ln}(T_{\rm lab})}{T_{\rm lab}} + \frac{0.3}{T_{\rm lab}^2}  \right )^{-1}   \label{steve1}\\
\sigma_{pp \rightarrow \pi^+ X} &=& \left ( 0.00717+0.0652 \frac{{\rm ln}(T_{\rm lab})}{T_{\rm lab}} + \frac{0.162}{T_{\rm lab}^2}  \right )^{-1}   \label{steve2} \\
\sigma_{pp \rightarrow \pi^- X} &=& \left ( 0.00456+ \frac{0.0846 }{T_{\rm lab}^{0.5}} + \frac{0.577}{T_{\rm lab}^{1.5}}  \right )^{-1}  \label{steve3}
\end{eqnarray}
where $T_{\rm lab}$ should be specified in GeV to give $\sigma$ in  mb.

The trouble with only having proton-proton (pp)  cross sections is that nuclei also contain neutrons, and so one needs cross sections for pion production from neutron-neutron (nn) and neutron-proton (np) collisions.  The {\em exclusive} reactions for single pion production  are listed below. 
\begin{eqnarray}
pp   &\rightarrow&   pp \pi^0    \hspace*{3.1cm} [\times 2] \\
   &\rightarrow&   pn \pi^+   \hspace*{3cm} [\times 2] 
\end{eqnarray}
and
\begin{eqnarray}
nn   &\rightarrow&   nn \pi^0    \hspace*{3.05cm} [\times 2] \\
  &\rightarrow&   np \pi^-   \hspace*{3cm} [\times 2] 
\end{eqnarray}
and 
 \begin{eqnarray}
pn  &\rightarrow&   pn \pi^0    \hspace*{3.1cm} [\times 2] \\
&\rightarrow&   nn \pi^+   \hspace*{3cm} [\times 2] \\
   &\rightarrow&   pp \pi^-   \hspace*{3.08cm} [\times 2] 
\end{eqnarray}
The number in square brackets after some reactions indicates that the reaction can proceed in a number of different ways and therefore the number of particles produced needs to be multiplied by the number in square brackets. For example the reaction $pp\rightarrow pn\pi^+$ can also proceed as
$pp\rightarrow np\pi^+$, with the pion being produced from the other nucleon.
Exclusive reactions for double pion production in terms of initial states are
 \begin{eqnarray}
pp &\rightarrow & pp \pi^0 \pi^0  \\
&\rightarrow & pp\pi^+\pi^- \hspace*{3cm} [\times 2] \\
&\rightarrow & pn\pi^+\pi^0 \hspace*{3.05cm} [\times 2] \\
&\rightarrow & nn\pi^+\pi^+
\end{eqnarray}
and
 \begin{eqnarray}
nn &\rightarrow &   nn\pi^0\pi^0\\
&\rightarrow & nn\pi^+\pi^- \hspace*{3cm} [\times 2] \\
&\rightarrow & pp\pi^-\pi^-\\
&\rightarrow & np\pi^-\pi^0 \hspace*{3.1cm} [\times 2] 
\end{eqnarray}
and
 \begin{eqnarray}
pn &\rightarrow & np\pi^0\pi^0\\
&\rightarrow & np\pi^+\pi^- \hspace*{3cm} [\times 2] \\
&\rightarrow & nn\pi^+\pi^0 \hspace*{3.05cm} [\times 2] \\
&\rightarrow & pp\pi^-\pi^0  \hspace*{3.1cm} [\times 2] 
\end{eqnarray}

By considering the reactions with $pp$ in the initial state, we see that the ratio of the numbers of pions is 
\begin{eqnarray}
\pi^+ : \pi^- : \pi^0 &=&  8:2:6 =4:1:3
\end{eqnarray}
Thus at {\em low energy} (i.e. around the two pion threshold) we expect the $pp$ cross sections to be in this ratio, namely
\begin{eqnarray}
\sigma_{pp\rightarrow \pi^+X} : \sigma_{pp\rightarrow \pi^-X}:\sigma_{pp\rightarrow \pi^0X} &=&  4:1:3  \hspace*{2cm} {\rm  (low \; energy)}  \label{ratio}
\end{eqnarray}
We therefore expect that  if we divide the $\sigma_{pp\rightarrow \pi^+X}$ by 4  and $\sigma_{pp\rightarrow \pi^0X}$ by 3  then the three pion cross sections should be roughly the same at low energy.  If one plots the cross sections this  is seen to be approximately true for low energy.

Thus we conclude that the $\pi^+$ to $\pi^-$ ratio for pp reactions at low energy is given by
\begin{eqnarray}
\frac{\sigma_{pp\rightarrow \pi^+X} }{ \sigma_{pp\rightarrow \pi^-X} }= 4  \hspace*{2cm} {\rm  (low \; energy)} 
\end{eqnarray}
A similar analysis leads us to conclude that the $\pi^+$ to $\pi^-$ ratio for nn reactions at low energy is given by
\begin{eqnarray}
\frac{\sigma_{nn\rightarrow \pi^+X} }{ \sigma_{nn\rightarrow \pi^-X} }= \frac{1}{4}  \hspace*{2cm} {\rm  (low \; energy)} 
\end{eqnarray} 
and  for pn reactions 
\begin{eqnarray}
\frac{\sigma_{pn\rightarrow \pi^+X} }{ \sigma_{pn\rightarrow \pi^-X} }= 1   \\   \nonumber   \\   \nonumber 
\end{eqnarray}

\section{Nucleus-Nucleus cross sections}
For  nucleus-nucleus collisions, denote the probability of a proton-proton reaction as
\begin{eqnarray}
{\rm Prob (pp) } = \frac{Z_P}{A_P} \frac{Z_T}{A_T}
\end{eqnarray}
where $Z$ and $A$ are the proton and nucleon numbers of the Projectile and Target. The probability of a  neutron-neutron reaction is 
\begin{eqnarray}
{\rm Prob (nn) } = \frac{A_P-Z_P}{A_P} \frac{A_T-Z_T}{A_T}
\end{eqnarray}
For nucleus-nucleus collisions at low energy we therefore expect the  $\pi^+$ to $\pi^-$ ratio to be
\begin{eqnarray}
\frac{\sigma_{AA\rightarrow \pi^+X} }{ \sigma_{AA\rightarrow \pi^-X} } &=&
{\rm Prob (pp) }  \times 4 \;\; +\;\;{\rm Prob (nn) }  \times \frac{1}{4} \hspace*{2cm} {\rm  (low \; energy)}  \\
&=&
\left ( \frac{Z_P}{A_P} \frac{Z_T}{A_T}\times 4 \right )  + \left (\frac{A_P-Z_P}{A_P} \frac{A_T-Z_T}{A_T} \times \frac{1}{4}  \right ) \hspace*{1cm} {\rm  (low \; energy)}  
\end{eqnarray}
Generalising to higher energy therefore gives
\begin{eqnarray}
\frac{\sigma_{AA\rightarrow \pi^+X} }{ \sigma_{AA\rightarrow \pi^-X} } &=&
\left ( \frac{Z_P}{A_P} \frac{Z_T}{A_T}   \frac{\sigma_{pp\rightarrow \pi^+X} }{ \sigma_{pp\rightarrow \pi^-X} }  \right )   +  \left (\frac{A_P-Z_P}{A_P} \frac{A_T-Z_T}{A_T}  \frac{\sigma_{pp\rightarrow \pi^-X} }{ \sigma_{pp\rightarrow \pi^+X} } \right )      \label{semifinal}
\end{eqnarray}
However this argument only gives us the {\em ratio}. We need a model to get one of the cross sections. 
We first note that $\pi^+$ production can be well described using
\begin{eqnarray}
\sigma_{AA\rightarrow \pi^+X} \approx   (A_PA_T)^{2/3}\sigma_{pp\rightarrow \pi^+X} \label{pp}
\end{eqnarray}
In order to obtain $\sigma_{AA\rightarrow \pi^- X} $ we proceed as follows. Substitute (\ref{pp}) into (\ref{semifinal}) to obtain
\begin{eqnarray}
\sigma_{AA\rightarrow \pi^- X}  = \frac{ (A_PA_T)^{2/3}\sigma_{pp\rightarrow \pi^+X}}
{\left ( \frac{Z_P}{A_P} \frac{Z_T}{A_T}   \frac{\sigma_{pp\rightarrow \pi^+X} }{ \sigma_{pp\rightarrow \pi^-X} }  \right )   +  \left (\frac{A_P-Z_P}{A_P} \frac{A_T-Z_T}{A_T}  \frac{\sigma_{pp\rightarrow \pi^-X} }{ \sigma_{pp\rightarrow \pi^+X} } \right )   }  
\label{pp2}
\end{eqnarray}
Equations (\ref{pp}) and (\ref{pp2}) represent  formulas for charged pion production from nucleus-nucleus collisions.
However it will be seen below that somewhat better fits to data are obtained by slightly changing the 2/3 power appearing in   (\ref{pp}) and (\ref{pp2}). Therefore the final formulas for pion production in proton-nucleus and nucleus-nucleus collisions are
\begin{eqnarray}
\sigma_{AA\rightarrow \pi^+X} =  (A_PA_T)^{N_\pm}\;\sigma_{pp\rightarrow \pi^+X} \label{finalplus}
\end{eqnarray}
\begin{eqnarray}
\sigma_{AA\rightarrow \pi^-X}  = \frac{ (A_PA_T)^{N_\pm}\;\sigma_{pp\rightarrow \pi^+X}}
{\left ( \frac{Z_P}{A_P} \frac{Z_T}{A_T}   \frac{\sigma_{pp\rightarrow \pi^+X} }{ \sigma_{pp\rightarrow \pi^-X} }  \right )   +  \left (\frac{A_P-Z_P}{A_P} \frac{A_T-Z_T}{A_T}  \frac{\sigma_{pp\rightarrow \pi^-X} }{ \sigma_{pp\rightarrow \pi^+X} } \right )   }  
\label{finalminus}
\end{eqnarray}
\begin{eqnarray}
\sigma_{AA\rightarrow \pi^0X}  =   (A_PA_T)^{N_0}\;\sigma_{pp\rightarrow \pi^0X} \label{finalzero}
\end{eqnarray}
where equations (\ref{steve1}, \ref{steve2}, \ref{steve3}) are used to obtain the elementary pion cross sections.  We have found that the best fit to the data is obtained with the values
\begin{eqnarray}
N_\pm &=& 2.2/3   \label{number1} \\
N_0 &=& 2.4/3   \label{number2}
\end{eqnarray}
Comparisons have been made between the theory developed above and the data from References \cite{nagamiya} and \cite{schwalb}. The comparisons can be seen in Figures 1 - 8.
 It is evident that agreement between theory and experiment is good.

\section{Conclusions}
The present work has investigated neutral and charged pion production in proton-nucleus and nucleus-nucleus collisions at the intermediate energy (i.e. in the GeV region)  relevant to space radiation. Total cross sections for inclusive processes have been calculated and compared to experiment. The  parameterizations  developed in equations 
 (\ref{finalplus}, \ref{finalminus}, \ref{finalzero}, \ref{number1}, \ref{number2}) are in good agreement with experiment.  For  charged pion  production in nucleus-nucleus reactions the theory overlaps all error bars from 0.4  to 2.1  AGeV (Figs. 1 - 5).  For neutral pion production in nucleus-nucleus reactions at 1 AGeV there are only 3 data points and the agreement is satisfactory (Fig. 6).  For proton-nucleus reactions, only $\pi^-$ data is available. The agreement between theory and experiment is satisfactory Figs. 7 - 8.
 The equations developed herein are in a form that is  suitable for immediate input into heavy-ion  transport codes. The formalism is expected to be valid  for total cross sections in the GeV region.

\begin{center}
{\bf Acknowledgements}
\end{center}

\noindent
This work was supported by NASA Research Grant NAG8-1901.

\newpage
\hspace*{1.5cm}\includegraphics[width=10.3cm]{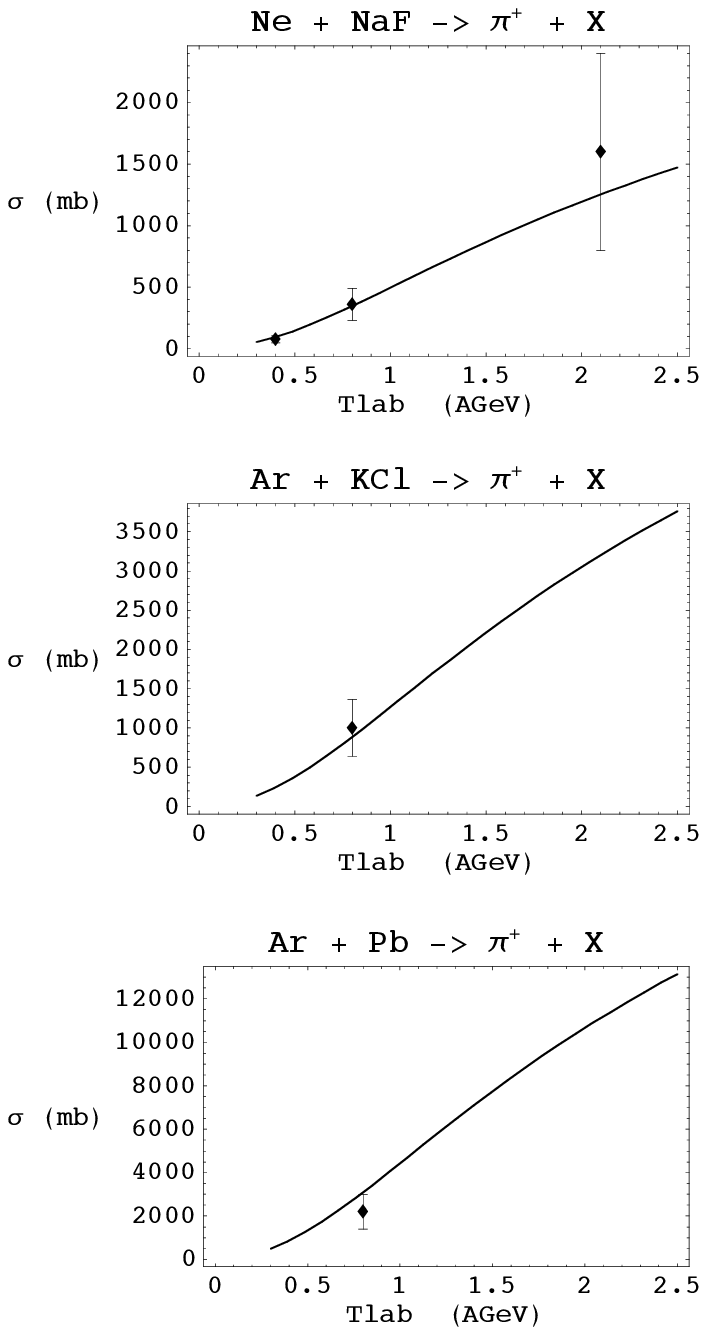}
\begin{quote}
 Figure 1.  Theory versus experiment for $\pi^+$ production in nucleus-nucleus collisions. The 
 line corresponds to equation (\ref{finalplus}).  Data are from reference \cite{nagamiya}.
\end{quote}

\newpage
\hspace*{1.5cm}\includegraphics[width=10.3cm]{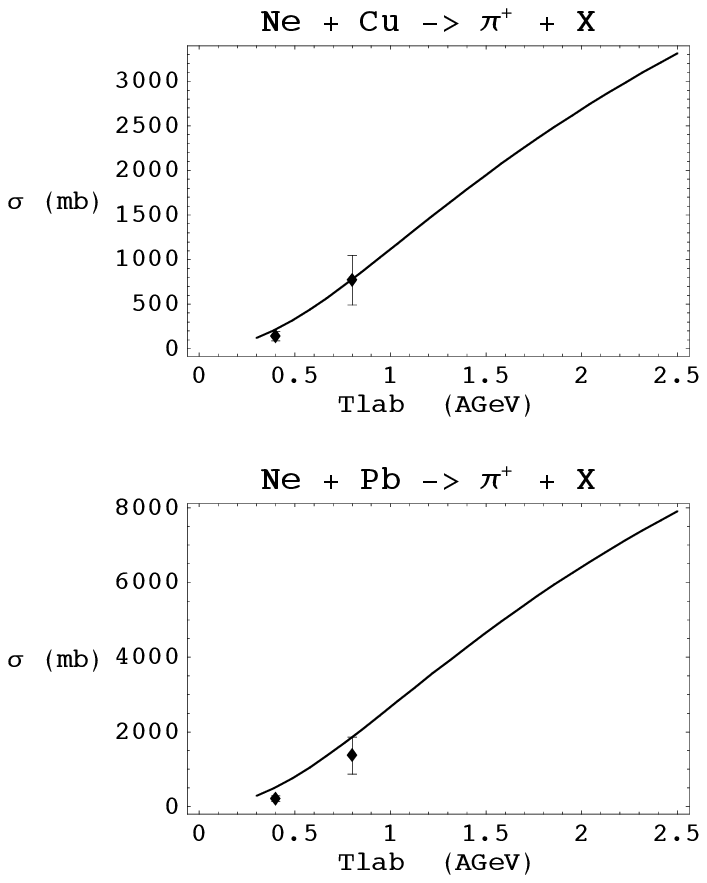}
\begin{quote}
 Figure 2.  Theory versus experiment for $\pi^+$ production in nucleus-nucleus collisions. The 
 line corresponds to equation  (\ref{finalplus}).  Data are from reference \cite{nagamiya}.
 \end{quote}

\newpage
\hspace*{1.5cm}\includegraphics[width=10.3cm]{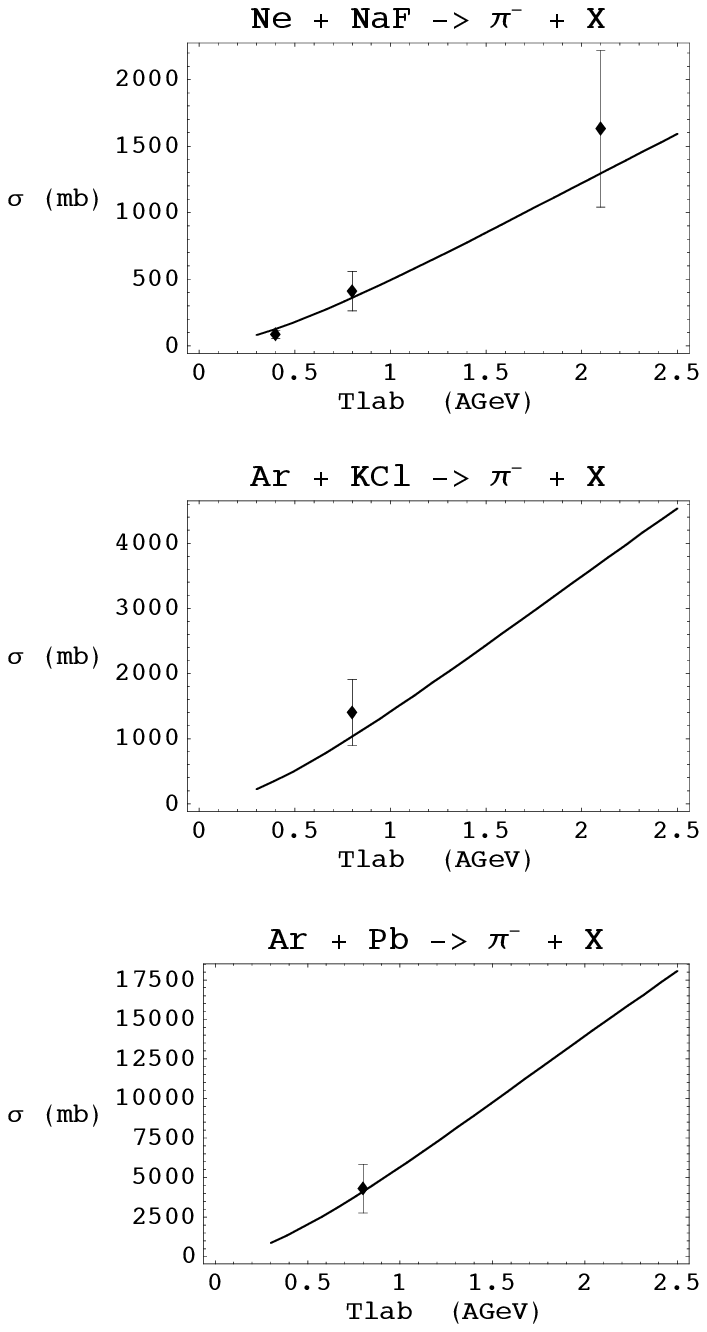}
\begin{quote}
 Figure 3.       Theory versus experiment for $\pi^-$ production in nucleus-nucleus collisions. The 
 line corresponds to equation  (\ref{finalminus}).  Data are from reference \cite{nagamiya}.
 \end{quote}

\newpage
\hspace*{1.5cm}\includegraphics[width=10.3cm]{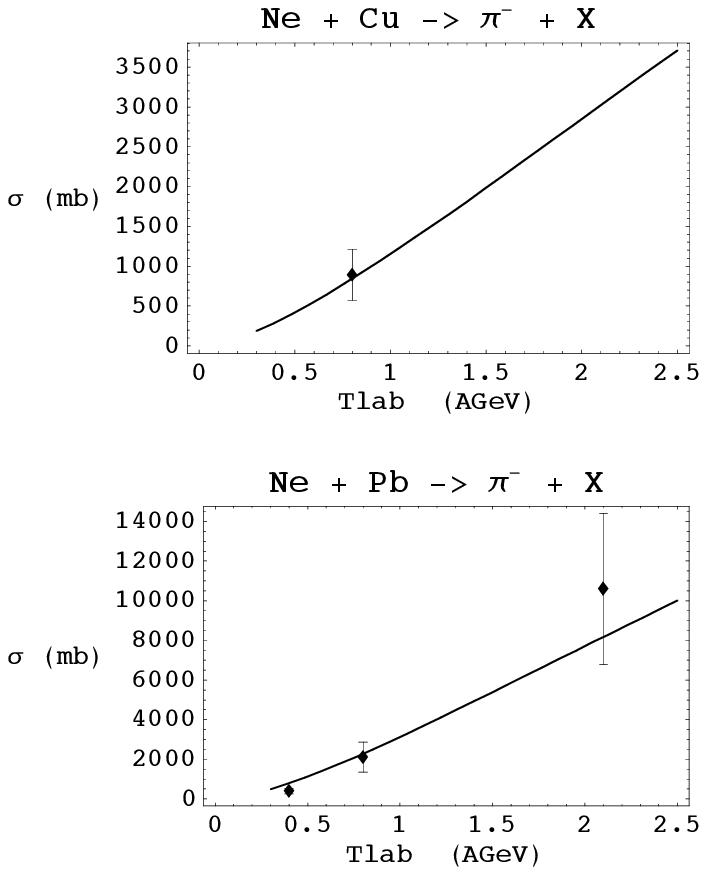}
\begin{quote}
 Figure 4.       Theory versus experiment for $\pi^-$ production in nucleus-nucleus collisions. The 
 line corresponds to equation  (\ref{finalminus}).  Data are from reference \cite{nagamiya}.
 \end{quote}

\newpage
\hspace*{1.5cm}\includegraphics[width=10.3cm]{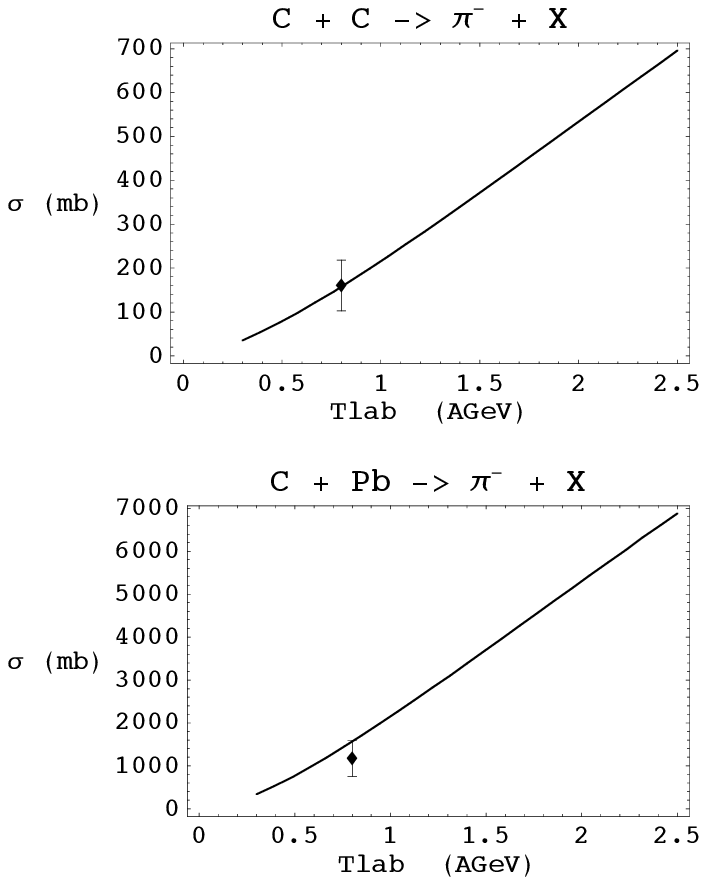}
\begin{quote}
 Figure 5.       Theory versus experiment for $\pi^-$ production in nucleus-nucleus collisions. The 
 line corresponds to equation  (\ref{finalminus}).  Data are from reference \cite{nagamiya}.
 \end{quote}

\newpage
\hspace*{1.5cm}\includegraphics[width=10.3cm]{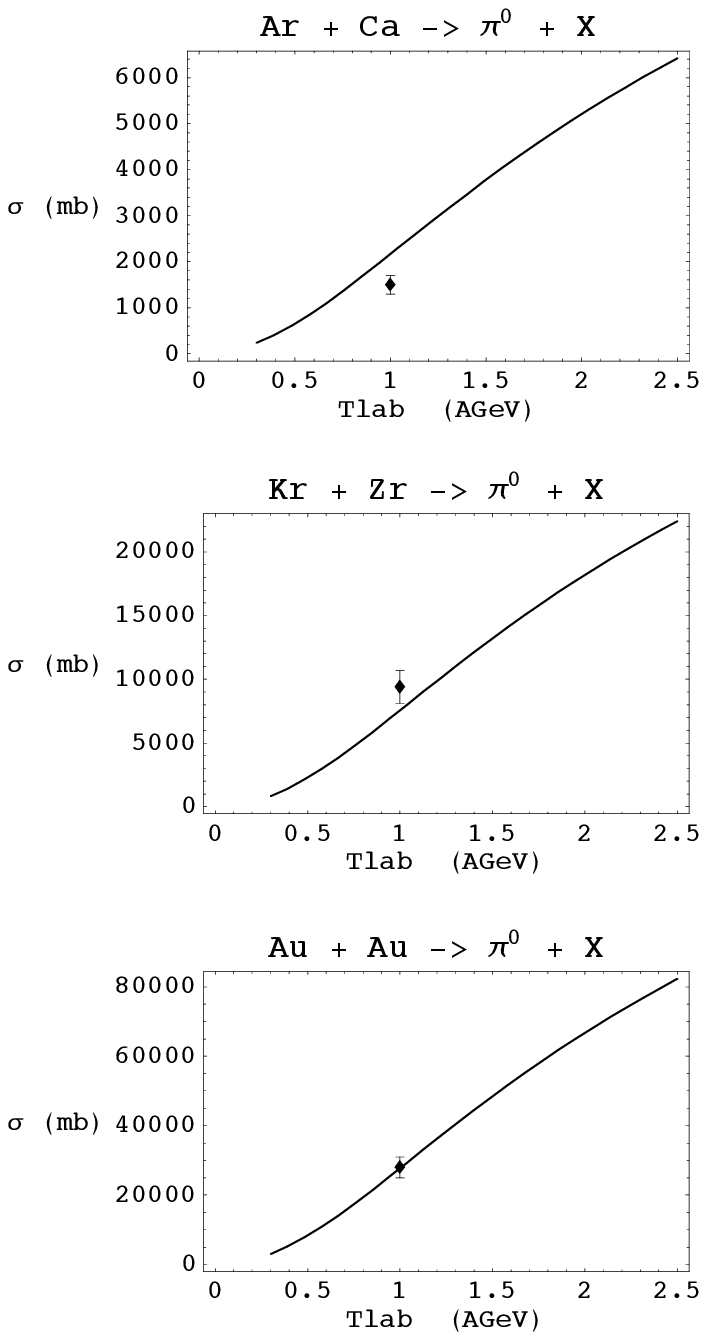}
\begin{quote}
 Figure 6.       Theory versus experiment for $\pi^0$ production in nucleus-nucleus collisions. The 
 line corresponds to equation  (\ref{finalzero}).  Data are from reference \cite{schwalb}.
 \end{quote}

\newpage
\hspace*{1.5cm}\includegraphics[width=10.1cm]{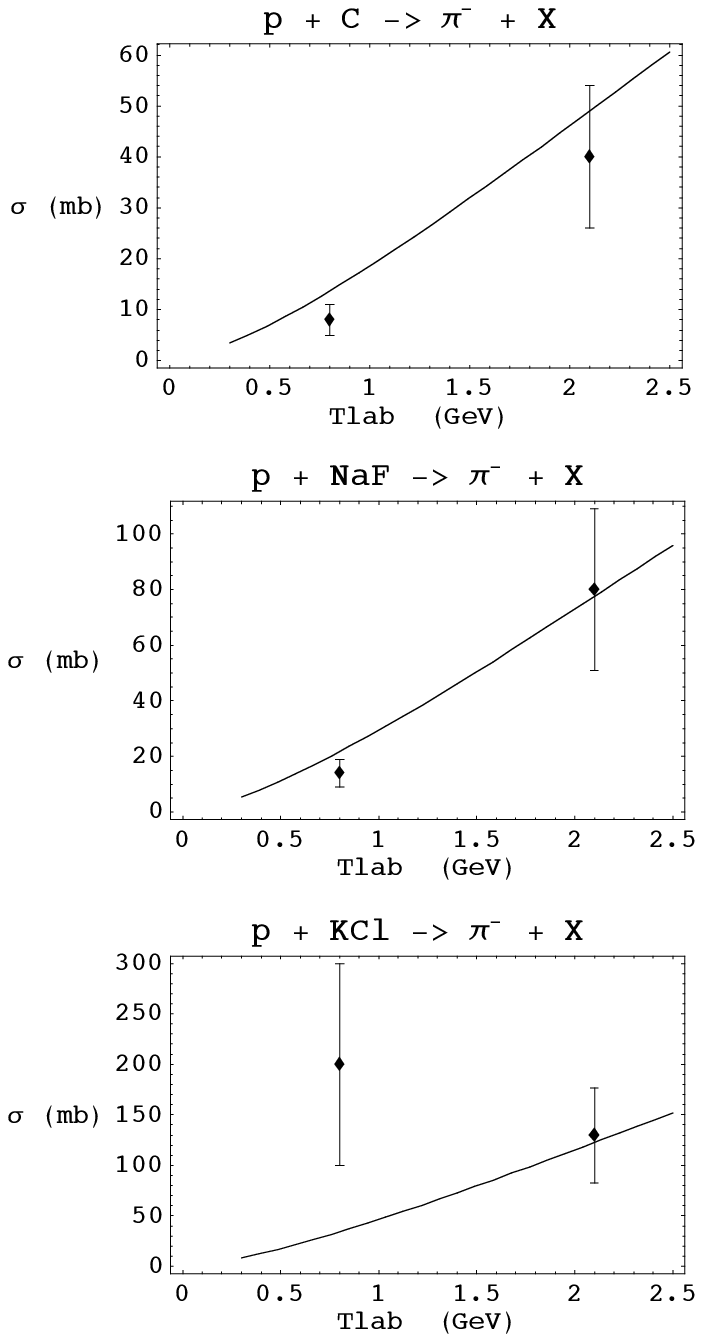}
\begin{quote}
 Figure 7.       Theory versus experiment for $\pi^-$ production in proton-nucleus collisions. The 
 line corresponds to equation  (\ref{finalminus}).  Data are from reference \cite{nagamiya}.
 \end{quote}

\newpage
\hspace*{1.5cm}\includegraphics[width=10.3cm]{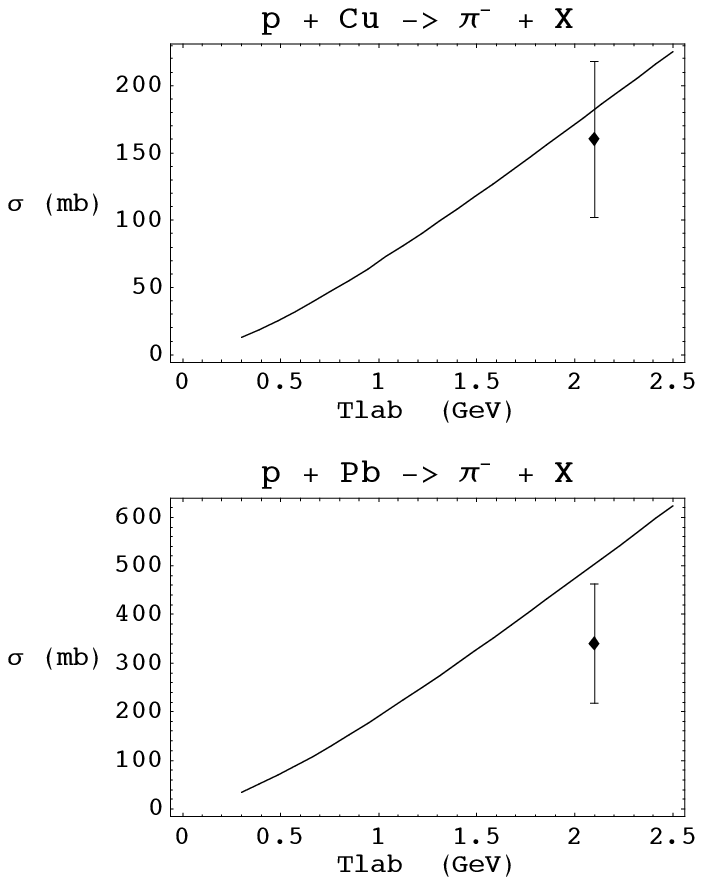}
\begin{quote}
 Figure 8.              Theory versus experiment for $\pi^-$ production in proton-nucleus collisions. The 
 line corresponds to equation  (\ref{finalminus}). Data are from reference \cite{nagamiya}.
 \end{quote}

\end{document}